\documentclass[journal]{IEEEtran}
\usepackage{cite}

\ifCLASSINFOpdf
\usepackage[pdftex]{graphicx}
\else
\fi
\usepackage{amsmath}
\usepackage{array}
\usepackage{amssymb}
\usepackage{booktabs,colortbl}
\usepackage{color}
\usepackage{eurosym}
\usepackage[font=small]{caption}
\usepackage{subcaption}
\usepackage{dsfont}
\usepackage{relsize}
\usepackage{mathtools}

\newcommand{\Dtk}{\delta}

\newcommand{\ds}{\{\Pgsch{k}\}_{k \in \KDiS}}

\newcommand{\CP}{\mathcal{P}}
\newcommand{\CE}{\mathcal{E}}

\newcommand{\ex}{\mathbb{E}}

\newcommand{\pload}{\bold{l}}
\newcommand{\eload}{e_{\text{l}}}
\newcommand{\plforz}{\plfor(k|\kzero)}
\newcommand{\plfor}{\rv{L}}
\newcommand{\dplfor}{\rv{\Delta L}}

\newcommand{\mun}{\mu_{\text{n}}}
\newcommand{\bmu}{\boldsymbol{\mu}}

\newcommand{\expPl}[1]{\hat{\bold{l}}(#1)}
\newcommand{\expPs}[1]{\hat{\bold{p}}_{0}(#1)}

\newcommand{\expPc}[1]{\hat{\bold{p}}(#1)}
\newcommand{\expEc}[1]{\hat{e}(#1)}
\newcommand{\psto}{\bold{p}_{0}}

\newcommand{\pagg}{\bold{p}}
\newcommand{\eagg}{e}
\newcommand{\rvpagg}{\rv{P}}
\newcommand{\rveagg}{\rv{E}}

\newcommand{\pvh}[1]{\bold{p}_{#1}}
\newcommand{\evh}[1]{e_{#1}}
\newcommand{\arrvh}[1]{k_{#1}^{\text{a}}} 
\newcommand{\depvh}[1]{k_{#1}^{\text{d}}} 
\newcommand{\evhmind}[1]{e_{#1}^{\text{min}}} 
\newcommand{\rvevh}[1]{\rv{E}_{#1}}

\newcommand{\Pgsch}[1]{\tilde{\bold{g}}(#1)}

\newcommand{\pgrid}{\bold{g}}
\newcommand{\dpg}{\boldsymbol{\sigma}}

\newcommand{\kstart}{k^{\text{b}}}

\newcommand{\kzero}{k^0}
\newcommand{\KDiS}{\mathcal{S}}

\newcommand{\Ivec}{\mathds{1}}
\newcommand{\Ivect}{\mathds{1}^{\top}}
\newcommand{\matrixproduct}{\bold{M}}

\newcommand{\rv}[1]{\mathsf{#1}}

\newcommand{\nonl}{\renewcommand{\nl}{\let\nl\oldnl}}

\newtheorem{rema}{Remark}

\DeclareCaptionFont{tiny}{\tiny}

\begin{document}
%
\title{Reliable Dispatch of Renewable Generation via Charging of Dynamic PEV Populations
}
%
%
%


\author{Riccardo Remo Appino,~\IEEEmembership{Member,~IEEE,}
        Miguel Mu\~noz-Ortiz,
        Jorge \'Angel Gonz\'alez Ordiano,
        Ralf Mikut, 
        Veit Hagenmeyer, 
        and~Timm Faulwasser,~\IEEEmembership{Member,~IEEE}
\thanks{All authors are with the Institute for Automation and Applied Informatics,
Karlsruhe Institute of Technology, 76344 Eggenstein-Leopoldshafen, Germany e-mail: riccardo.appino@kit.edu.}
\thanks{Manuscript received May 8, 2018; revised .}}

%
%

\markboth{IEEE Transaction on Power Systems}
{Shell \MakeLowercase{\textit{et al.}}: Bare Demo of IEEEtran.cls for IEEE Journals}
%



\maketitle

\begin{abstract}
The inherent storage of plug-in electric vehicles is likely to foster the integration  of intermittent generation from renewable energy sources into existing power systems.
In the present paper, we propose a three-stage scheme to the end of achieving dispatchability of a system composed of plug-in electric vehicles and intermittent generation.
The main difficulties in dispatching such a system are the uncertainties inherent to intermittent generation and the time-varying aggregation of vehicles.
We propose to address the former by means of probabilistic forecasts and we approach the latter with separate stage-specific models. 
Specifically, we first compute a dispatch schedule, using probabilistic forecasts together with an aggregated dynamic model of the system.  
The power output of the single devices are set subsequently, using deterministic forecasts and device-specific models. 
We draw upon a simulation study based on real data of generation and vehicle traffic to validate our findings.
\end{abstract}


%
\IEEEpeerreviewmaketitle

\section{Introduction}
\label{sec:intro}

Renewable generation from wind turbines and photovoltaics has the potential to enhance energy sustainability.
However, its penetration in the existing electric utility systems causes problems because of its intermittent nature, cf. \cite{Denholm07}.
Thus, additional resources are required, mainly capacity reserves and storage. 
Plug-In Electric Vehicles (PEVs) are a promising technology in this context: vehicles inherently rely on storage for their transportation function and they are on average in-use only for $4\%$ of the time \cite{Kempton05}. 
To exploit this storage for system operation, several ``smart" charging strategies for PEVs have been proposed,   
see \cite{Mukherjee15} for a recent review.
In the present paper, we investigate scheduling and control to the end of achieving dispatchability of a grid-connected system composed of intermittent generation, a PEV charging station, and an additional storage system.
In doing so we will consider two crucial aspects: uncertainties and time-varying device aggregation due to vehicles arriving and leaving.

Uncertainties enter the problem in terms of forecasts of intermittent generation and in terms of driving habits. 
Most works in the literature address uncertainty using scenario-based optimization, e.g. \cite{Momber15}.
Other works, instead, approach uncertainty by means of probabilistic forecast and chance constraint optimization, e.g. \cite{Kou16}.
In \cite{Appino18b}, it is shown that chance constrained optimization, based on probabilistic forecasts, might outperform scenario-based schemes.

Device aggregation is the second crucial aspect. 
Coordination of a population of direct controllable loads via one common cluster manager is well discussed, e.g. \cite{Mathieu13,Bernstein15a,Subramanian13}. 
These works show that managing a group of collaborating devices as one entity not only simplifies the control of the system but also reduces the need of reserves to compensate for uncertainties. 

In the present paper we aim at dispatchability of a charging station for PEVs, combining chance constraint optimization and device aggregation.
Our major contribution is the extension of the scheduling and control scheme described in \cite{Appino18a,Appino18b}, which cannot be applied as-it-is to systems including more than a single static energy storage.
The novel method described in this paper is instead able to cope with a dynamic aggregation of energy storage and direct controllable loads, considering multiple sources of uncertainty (e.g.: forecast errors and unknown state of charge of PEVs upon arrival).

The remainder of the present paper is organized as follows: Section \ref{sec:problem_state} covers the problem setup; Section \ref{sec:3stage_sched} presents the developed approach; Section \ref{sec:sim_and_results} details simulation results; Section \ref{sec:conclusion} summarizes our findings.

\section*{Notation}

Bold letters $\bold{x}\in \mathbb{R}^+_0 \times \mathbb{R}^-_0  \subset {\mathbb{R}}^2$ denote power flows. The two dimensions of these vectors discriminate between power flow directions. 
The set $\mathcal{K} \subset \mathbb{N}$ denotes a time period, divided into time steps $k \in \mathcal{K}$ of equal duration $\Dtk$. 
The variable $\bold{x}(k)$ refers to the average power flow over the $k$-th step.
Given a random variable $\rv{X}$, we indicate with $f_{\rv{X}}(\cdot)$ its Probability Distribution Function (PDF) and $F_{\rv{X}}(\cdot)$ is its Cumulative Density Function (CDF).
Finally, we use $\Ivec := \begin{bmatrix} 1 & 1 \end{bmatrix}^{\top}$.

\section{Problem Statement}
\label{sec:problem_state}

\subsection{System Description}
\label{sec:system_description}

\begin{figure}
\vspace{-0.2cm}
\centering
\includegraphics[width=0.42\textwidth]{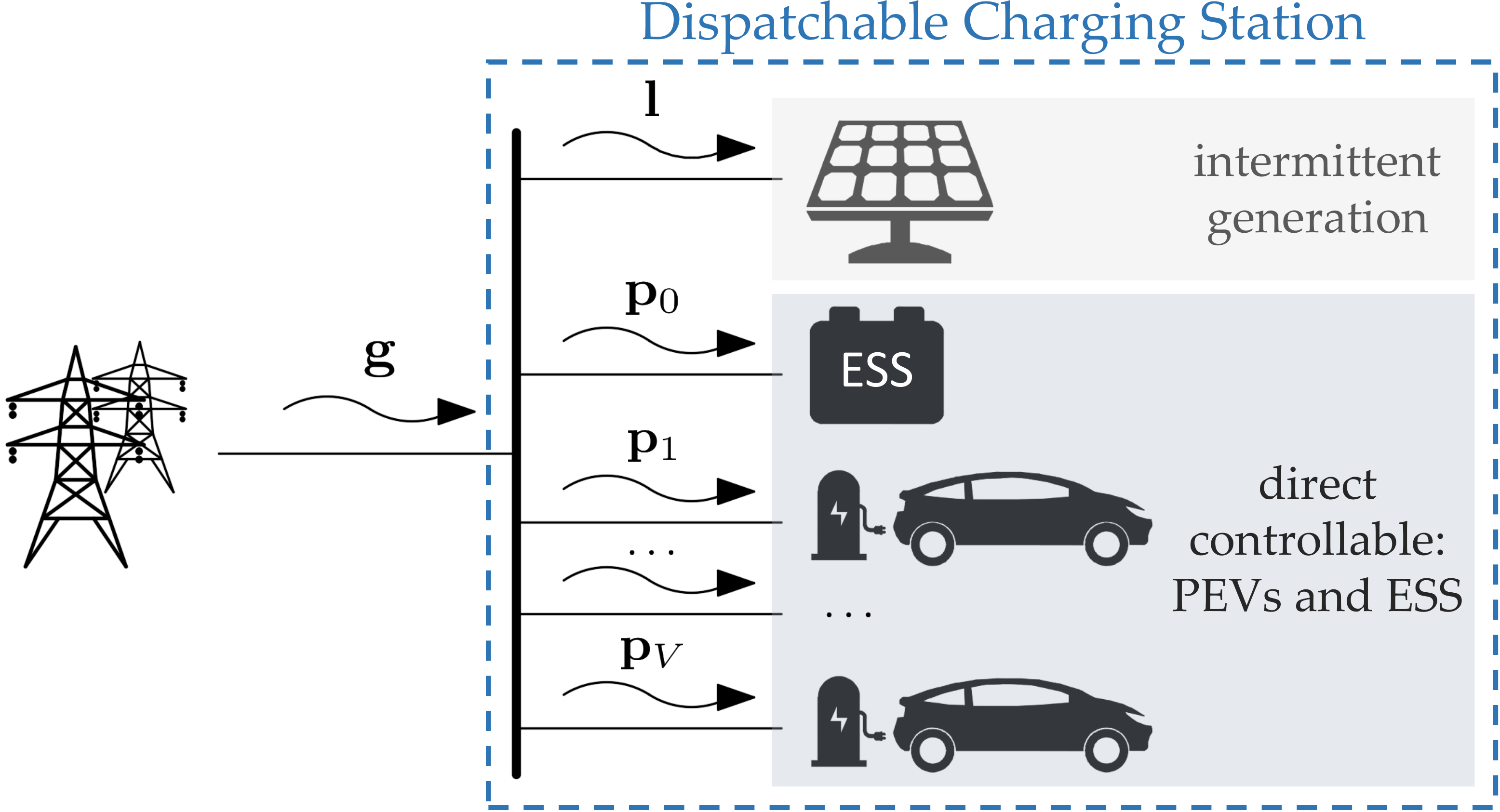}
\caption{Schematic diagram of a \textit{dispatchable charging station}. \label{fig:Scheme_DCS} 
}
\vspace{-0.5cm}
\end{figure}

This paper addresses scheduling and operation of a system composed of a charging station for Plug-in Electric Vehicles (PEVs), an Energy Storage System (ESS) and a generator exploiting intermittent renewable energy sources, such as a PhotoVoltaic (PV) generator. 
All the devices are connected to the same bus, as schematically represented in Fig. \ref{fig:Scheme_DCS}.
The connections are assumed lossless and such that the system components are able to exchange power mutually without any technical limit.
Note that the number of PEVs might change in time.  
Given a time interval $\mathcal{K}$, the set $\mathcal{V}_{\mathcal{K}} = [1 \hdots V] \in \mathbb{N}$ collects all the indices associated with the PEVs connecting to the charging station for at least one $k\in \mathcal{K}$ and the set $\mathcal{V}_{\mathcal{K}}(k) \subset\mathcal{V}_{\mathcal{K}}$ contains the indexes of the PEVs connected at $k$.
The active power balance is 
\begin{equation}
\label{eq:power_balance_general}
\Ivect \pgrid(k) = \Ivect \pload(k) + \sum_{\mathclap{j\in \mathcal{V}^0_{\mathcal{K}}(k)}} \Ivect \pvh{j}(k).
\end{equation}
Hereby, $\pgrid(k) = \begin{bmatrix} g^-(k) & g^+(k)\end{bmatrix}^\top$ denotes the power exchange with the main grid, $\pload(k)$ the aggregated power output from uncontrollable generation, and $\pvh{j}(k)$ the power output of the $j$-th controllable device. 
The sign convention is the same for all power variables: the first element of each vector denotes a power flow directed as depicted in Fig. \ref{fig:Scheme_DCS}, while the second element indicates power flowing in the opposite direction. 
Indices $j$ are collected in the set $\mathcal{V}^0_{\mathcal{K}}(k)=\{0,\mathcal{V}_{\mathcal{K}}(k)\}$ with $0$ indicating the ESS.

Each  $\pvh{j}(k)$ influences the State Of Charge (SOC) of its corresponding device, $\evh{j}(k)$.
This state follows
\begin{equation}
\label{eq:complete_pev_din_det}
\evh{j}(k+1) = \evh{j}(k) + \bmu^\top \pvh{j}(k) \Dtk, 
\end{equation}
where $\bmu = \begin{bmatrix}
(1 - \mun) & (1 + \mun)
\end{bmatrix}^{\top}$. 
The conversion losses are modeled by a constant coefficient $\mun \in [0,1]$, exploiting the discrimination between the different directions of $\pvh{j}$.
For the sake of readability, we assume in the following that the ESS and all the PEVs are subject to the same efficiency $\mun$.
Notice that the individual PEVs are not constantly connected to the charging station.
Therefore $\evh{v}(k)$, $v \in \mathcal{V}_{\mathcal{K}}$, exists only within the interval $[\arrvh{v}, \depvh{v}]$, where $\arrvh{v}$ and $\depvh{v}$ denote the arrival and departure times of the $v$-th vehicle, respectively.
Furthermore, we consider the ordered index set $\mathcal{A}_{\mathcal{K}}(k) = \{v \in \mathcal{V}_{\mathcal{K}} | \arrvh{v} \leq k \}$, representing the vehicles arriving until $k$, and the set $\mathcal{D}_{\KDiS}(k) = \{v \in \mathcal{V}_{\KDiS} |  \depvh{v} \leq k \}$, representing the vehicles leaving until $k$. 

Capacity and capability constraints limit the power output and the energy state of each device.
For a device $j \in  \mathcal{V}_{\mathcal{K}}^0$
\begin{subequations}
\label{eq:det_ess_limit}
\begin{gather}
\underline{p}_j \leq \Ivect \bold{p}_j(k) \leq \overline{p}_j, \label{eq:det_pow_limit}\\
\underline{e}_j \leq {e}_j(k) \leq \overline{e}_j, \label{eq:det_en_limit}
\end{gather}
\end{subequations}
where $\underline{p}_j$ and $\overline{p}_j$ are the minimum and maximum power output; $\underline{e}_j$ and $\overline{e}_j$ are the minimum and maximum energy capacity.

\subsection{Requirements}
 
The scheme proposed in the present paper aims at \emph{dispatch-as-scheduled} of the power exchange $\pgrid$, cf. Fig. \ref{fig:Scheme_DCS}, by means of a coordinated and collaborative control of the system components.
Thus, we refer to the system under analysis as \textit{Dispatchable Charging Station} (DCS), motivated by
the concept of \text{dispatchable feeder} first introduced in \cite{Sossan16a}.
Similar to \cite{Sossan16a}, we consider a scenario where the power exchange with the utility grid has to be regulated according to a pre-computed \textit{Dispatch Schedule} (DiS).
Operating revenues (or costs) can be directly associated to the DiS through a known cost function.
Deviations from the DiS, here called \textit{imbalances}, are limited by regulations or penalized in operation.
Furthermore, similar to \cite{Kou16}, we consider that the power exchanged with the PEVs does not directly relate to any operating revenue.
However, the DCS should still satisfy the charging requests of each PEV.
Specifically, a minimum SOC $\evhmind{v}$ should be achieved before departure, at $\depvh{v}$.  
Both $\evhmind{v}$ and $\depvh{v}$ are declared by the PEV upon arrival, assuming that $\evhmind{v}$ is reachable (feasible) at $\depvh{v}$ given power constraint \eqref{eq:det_pow_limit} and the initial SOC.
The selected scenario translates in the following requirements.

\subsubsection{Requirements on the dispatch schedule}
\label{sec:req_disp_sch}
The DiS extends over an interval $\KDiS= [\kstart, \kstart+S] \in \mathbb{N}$ and has to be computed before its application at $\kzero < \kstart$. 
We denote the DiS at $k$ with $\Pgsch{k}$ and the sequence of $\Pgsch{k}$ over the scheduling horizon with the shorthand notation $\ds$, i.e. 
${\ds=\{\Pgsch{\kstart},\Pgsch{\kstart+1},...,\Pgsch{\kstart+S}\}}$.
The DiS should be reliable and efficient. 
With reliable, we intend that the DCS should track the DiS---in operation---with at least a given probability, $(1-\varepsilon)$.
Throughout the paper, we refer to the probability $(1-\varepsilon)$ as \textit{reliability level}. 
Besides, the DiS should also be efficient: while the DiS should satisfy the reliability requirement, it should also best comply to other desired targets, such as peak shaving, price-based load shifting, and maximization of self-consumption.
These further requirements can be translated into a cost function of the DiS itself. 
Hence, we consider      
\begin{align}
\label{eq:cost_f}
C(\pgrid(k)) = &\pgrid^{\top}(k) \bold{C}(k) \pgrid(k)  + \bold{c}^{\top}(k) \pgrid(k) \nonumber \\ &\quad + c^\text{i} \Ivect (\pgrid(k)-\pgrid(k-1))^2, 
\end{align} 
where the elements of $\bold{C}(k)\in \mathbb{R}^{2\times2}$ and $\bold{c}^{\top}(k) \in \mathbb{R}^{2}$ are (known) time-varying cost coefficients, and $c^\text{i}$ is a constant weight penalizing the incremental change of $\pgrid$ (approximating the derivative of the DiS).
Peak shaving, price-based load shifting, and maximization of self-consumption are all scheduling requirements that can be described with \eqref{eq:cost_f}.

\begin{figure*}
\vspace{-0.2cm}
\centering
\includegraphics[width=0.99\textwidth]{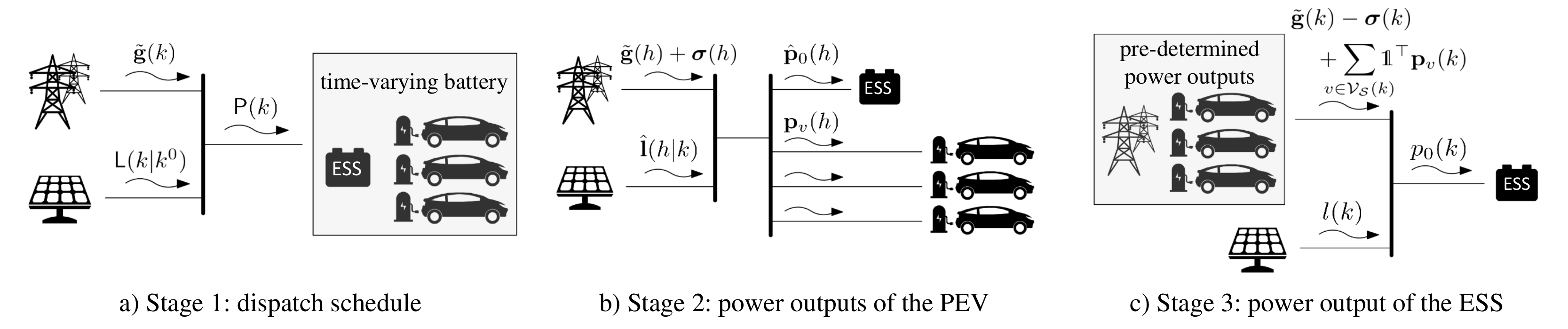}
\caption{Device aggregation at different stages \label{fig:scheduling_steps} 
}
\vspace{-0.5cm}
\end{figure*}

\subsubsection{Operation requirements}
\label{sec:operating_requirements}
During operation, the controller of the DCS should regulate the power output of the ESS and of the PEVs such that:
i) the power exchange $\pgrid(k)$ follows $\Pgsch{k}$ as close as possible, and ii)  upon departure the SOC of each $v$-th PEV satisfies its corresponding charging request,
\begin{equation}
\label{eq:minimum_SOC_PEV}
\evhmind{v} \leq \evh{v}(\depvh{v}) \leq \overline{e}_{v}.
\end{equation}
Observe that charging the battery of the $v$-th PEV beyond $\evhmind{v}$ does not yield any additional revenue, with the only exception (practically rare) of negative costs for purchase of power from the grid.

\section{Three-stage Scheduling and Operation} 
\label{sec:3stage_sched}

Similarly to \cite{Vandael13}, we partition the scheduling and operation of a DCS in \textit{three} subsequent decision stages. 
A separate optimization is conducted at each stage, depending on a stage-specific model. 
First, a DiS is computed based on an aggregation of the constraints of all controllable devices. 
Then, a charging plan for the PEVs is computed.
Finally, the power output of the ESS is regulated in compliance with the DiS. 
The various stages are depicted schematically in Fig. \ref{fig:scheduling_steps} and detailed in the following Subsections.   

\subsection{Stage 1: Dispatch Schedule}
\label{sec:stage1}
The computation of the DiS $\ds$ at $k_0$ constitutes the first decision stage. 

\subsubsection{Aggregated Model for Controllable Devices}
\label{sec:aggregated_model_descr}
With respect to the controllable devices---and extending the concept of ``time-varying battery" \cite{Mathieu13} or the similar schemes  \cite{Wenzel17, Vandael13}---we employ a model which aggregates both the ESS and the PEVs as in Fig. \ref{fig:scheduling_steps}a. 
This way, not only we reduce the complexity of the problem \cite{Mathieu13}, but we also reduce the requirement of reserves to cope with uncertainty \cite{Subramanian13}. 
In fact, the aggregation of the various devices into a ``time-varying battery" corresponds to an indistinct use of both the ESS and the PEVs to compensate for the uncertain power output. That is, in the computation of the DiS we ensure solely that the power required to maintain $\pgrid(k)$ at $\Pgsch{k}$ will be provided by some devices while leaving the decision of \textit{which} specific device will practically provide this power to a subsequent moment.
Therefore, the flexibility provided by the controllable devices is exploited to a greater extend.

\begin{figure}
\vspace{0.05cm}
\centering
\includegraphics[width=0.42\textwidth]{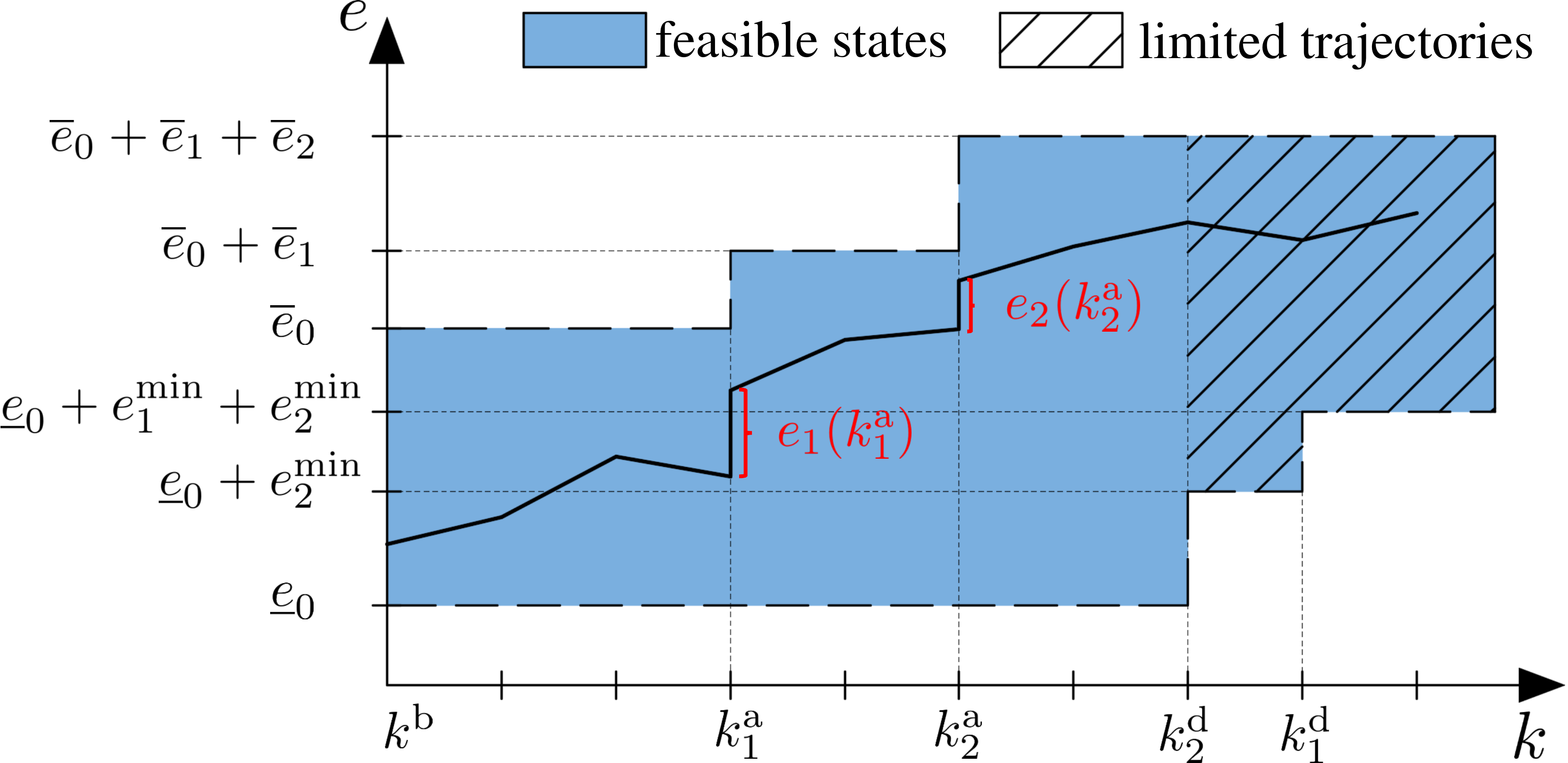}
\caption{Example of ``time-varying battery" capacity limits and evolution of $e(k)$ with $\underline{e}_{1}=\underline{e}_{2}=0$.\label{fig:Energy boundaries} 
}
\vspace{-0.5cm}
\end{figure}

The ``time-varying battery" is characterized by the power output $\pagg(k)$ and the SOC $\eagg(k)$ evolving according to 
\begin{equation}
\eagg(k+1) = \eagg(k) + \sum_{\mathclap{v\in \mathcal{A}_{\KDiS}(k)}} \evh{v}(\arrvh{v}) + \bmu^\top \pagg(k) \Dtk, 
\label{eq:complete_tvb_din_det}
\end{equation}
This behavior is similar to \eqref{eq:complete_pev_din_det}, with the addition of sudden energy increase at $\arrvh{j}$.\footnote{Note that the assumption of equal $\mun$ for all the connected devices can be dropped here by using an appropriate average of the various $\mun$.} 
In other words, when a PEV arrives, the energy stored in its battery is instantaneously added to the energy stored in the aggregated system. 
The capability and capacity constraints of the ``time-varying battery" are
\begin{gather}
\underline{p}(k) \leq \Ivect \pagg(k) \leq \overline{p}(k), \label{eq:det_pow_limit_aggregated}\\
\underline{e}(k) \leq \eagg(k) \leq \overline{e}(k). \label{eq:det_en_limit_aggregated}
\end{gather}
These limits depend on the presence of PEVs plugged to the DCS, thus are time-varying. 
In power constraint \eqref{eq:det_pow_limit_aggregated}, the power limits are given by the sum of the power limits of the devices connected to the DCS at $k$,
\begin{equation*}
\underline{p}(k) = \sum_{\mathclap{j\in \mathcal{V}^0_{\KDiS}(k)}} \underline{p}_{j}, \quad 
\overline{p}(k) = \sum_{\mathclap{j\in \mathcal{V}^0_{\KDiS}(k)}} \overline{p}_{j}.
\end{equation*}
For the energy constraint, instead, we use a different model: 
\begin{subequations}
\label{eq:aggregated_model_capacity_limits}
\begin{align}
\overline{e}(k) &= \overline{e}(k-1) + \sum_{\mathclap{v\in \mathcal{A}_{\KDiS}(k)}} \overline{e}_{v},\\
\underline{e}(k) &= \underline{e}(k-1) + \sum_{\mathclap{v\in \mathcal{A}_{\KDiS}(k)}} \underline{e}_{v} + \sum_{\mathclap{v\in \mathcal{D}_{\KDiS}(k)}} \evhmind{v}, \label{eq:aggregated_model_capacity_limits_lower}
\end{align}
\end{subequations}
with 
\begin{align*}
\overline{e}_{\text{c}}(\kzero) = \overline{e}_{0} + \sum_{\mathclap{v\in \mathcal{A}_{\KDiS}(\kzero)}} \overline{e}_{v}, \quad
\underline{e}_{\text{c}}(\kzero) = \underline{e}_{0} +  \sum_{\mathclap{v\in \mathcal{D}_{\KDiS}(\kzero)}} \underline{e}_{v}.
\end{align*}
An example of the energy constraint of a ``time-varying battery" is depicted in Fig. \ref{fig:Energy boundaries}. 

Note that the arrival of a PEV has a very different model compared to the departure of the same PEV. 
The arrival of the PEV corresponds to an increment of both lower and upper capacity limits \eqref{eq:aggregated_model_capacity_limits} together with an ``energy addition" of $\evh{v}(\arrvh{v})$ \eqref{eq:complete_tvb_din_det}.  
On the contrary, the departure of a vehicle is modeled only by an increment in the lower capacity limit, see \eqref{eq:aggregated_model_capacity_limits_lower}, without any change in the state of charge $\eagg$. 
The reason behind this choice is that using for the departure of a PEV the same scheme used for its arrival---which would correspond to a reduction of the boundaries $\overline{e}(\depvh{v})$ and $\underline{e}(\depvh{v})$ at $\depvh{v}$ and to a subtraction of $\evh{v}(\depvh{v})$ from $\eagg(\depvh{v}-1)$---would require the precise knowledge of $\evh{v}(\depvh{v})$.
However, the aggregated model does not distinguish among the SOC of its single components: once a PEV arrives at the DCS, its available storage capacity is added to the ``time-varying battery" and its state of charge becomes indistinguishable from the one of the other devices.
The asymmetric model employed in this work solves this issue, yet introducing an additional challenge.
In fact, once the $v$-th PEV leaves the DCS, the eventual excess of energy ${\evh{v}(\depvh{v})- \evhmind{v}}$ is not stored anymore within the devices connected to the DCS but it is still included in the aggregated model $\eagg$.
Consequently, not every trajectory connecting feasible states $e$ is actually a feasible trajectory, see Fig. \ref{fig:Energy boundaries}.  
We will discuss the implication of this limitation in Section \ref{sec:scheduling_algorithm_fst}.

\subsubsection{Sources of Uncertainty at the First Stage}
\label{sec:sources_of_uncertainty}
The power output from uncontrolled generation $\{\pload(k)\}_{k\in \KDiS}$ is yet unknown at $k_0$.
However, a probabilistic forecast for $\{\pload(k)\}_{k\in \KDiS}$ can be calculated \cite{GonzalezOrdiano17}. 
Hence, the inflexible power output at the first stage is represented by a random variable, 
$\plforz$, whose realization is $l(k)=\Ivect \pload(k)$.
\footnote{The forecasts provide knowledge on the quantiles of $\plforz$, where $k$ refers to the forecast horizon and $\kzero$ to the time when the forecast is conducted.
The CDF and the PDF of $\plforz$ could be obtained starting from these quantiles.} 
In computing the DiS, we consider that the ``time-varying battery" compensates---in operation---for the uncertainty affecting $\plforz$, such that the power exchange with the grid equals (in principle) the pre-computed schedule $\Pgsch{k}$ regardless of the realization of $\plfor(k)$.
This requirement leads to the probabilistic power balance of the system
\begin{equation}
\rvpagg(k) = \Ivect \Pgsch{k} - \plforz, \label{eq:gen_power_bal_1stage}
\end{equation}   
where the random variable $\rvpagg(k)$ describes the output of the ``time-varying battery".
The concept of this probabilistic power balance is similar to the one of probabilistic power flow \cite{Borkowska74}, meaning that, given a realization $l(k)$ of $\plforz$, the realization $p(k)$ of $\rvpagg(k)$ is such that
\begin{equation}
p(k) = \Ivect \Pgsch{k} - l(k). \label{eq:gen_power_bal_1stage_realization}
\end{equation}

Additional uncertainties can arise from the parameters of the PEVs \cite{Lee12}.
In the following, we assume that many characteristics of each $v$-th PEV---$\overline{p}_{v}, \underline{p}_{v}, \overline{e}_{v}, \underline{e}_{v}, \evhmind{v}, \arrvh{v}$, and $\depvh{v}$---are known at $\kzero$.\footnote{This assumption might seem unrealistic but it holds in different applications.
A trivial example is given by scenarios in which the PEVs are requested to declare their intended arrival and departure times on the day before.
A second case can be found in parking lots of office buildings, where the variance on $\arrvh{v}$, and $\depvh{v}$ is generally smaller than an hour \cite{Sarabi16}.
With respect to the DCS, if the variance on $\arrvh{v}$ and $\depvh{v}$ is of the same order of the duration of the dispatch intervals $\Dtk$, the arrival and departure times of a PEV could be considered as certain. }
In contrast, we consider that the SOC of the $v$-th PEV at $\arrvh{v}$ is uncertain and we describe it using a random variable $\rvevh{v}(\arrvh{v})$, with realization $\evh{v}(\arrvh{v})$.

From \eqref{eq:complete_tvb_din_det}, it follows that both uncertainties, i.e. $\rvpagg(k)$ and $\rvevh{v}(\arrvh{v})$, affect the SOC of the ``time varying-battery", which is therefore a random variable itself: $\rveagg(k)$ with realization $\eagg(k)$.
We describe the dynamics of $\rveagg(k)$ using a stochastic model similar to the one in \cite{Appino18b}, where the conversion losses of the ``time-varying battery" are approximated to their expected value. 
However, such a model is extended in the following to include the additional uncertainty $\rvevh{v}(\arrvh{v})$. 

\subsubsection{Stochastic Model of Stored Energy}
\label{sec:stoch_model_energy}

Similar to \cite{Appino18b}, we divide deterministic and stochastic variables as follows
\begin{subequations}
\label{eq:exp_value_notation}
\begin{align}
\Ivect \expPl{k|\kzero} &= \ex[\plforz], \\
\plforz &= \Ivect \expPl{k|\kzero} + \dplfor(k),
\end{align}
\begin{align}
\rvpagg(k) &= \Ivect \expPc{k} + \rv{\Delta P}(k), \\
\rveagg(k) &= \expEc{k_0} + \rv{\Delta E}(k),
\end{align}
\end{subequations}
and impose
\begin{equation}
\label{eq:exp_ess_power}
\Ivect \expPc{k} = \Ivect \Pgsch{k} - \Ivect \expPl{k|\kzero}.
\end{equation}
Consequently, it holds that
\begin{equation}
\rv{\Delta P}(k) = - \rv{\Delta L}(k). \label{eq:deviation_p_bal}
\end{equation}

Excluding the ``energy increment" due to vehicles arrival , the deterministic part of $\rveagg(k)$ evolves in according to \eqref{eq:complete_tvb_din_det}
\begin{equation}
\label{eq:deterministic_ec-dynamics}
\expEc{k+1} = \expEc{k} + \bmu^\top \expPc{k} \Dtk. 
\end{equation}

\begin{figure}
	\centering
        \includegraphics[width=0.46\textwidth]{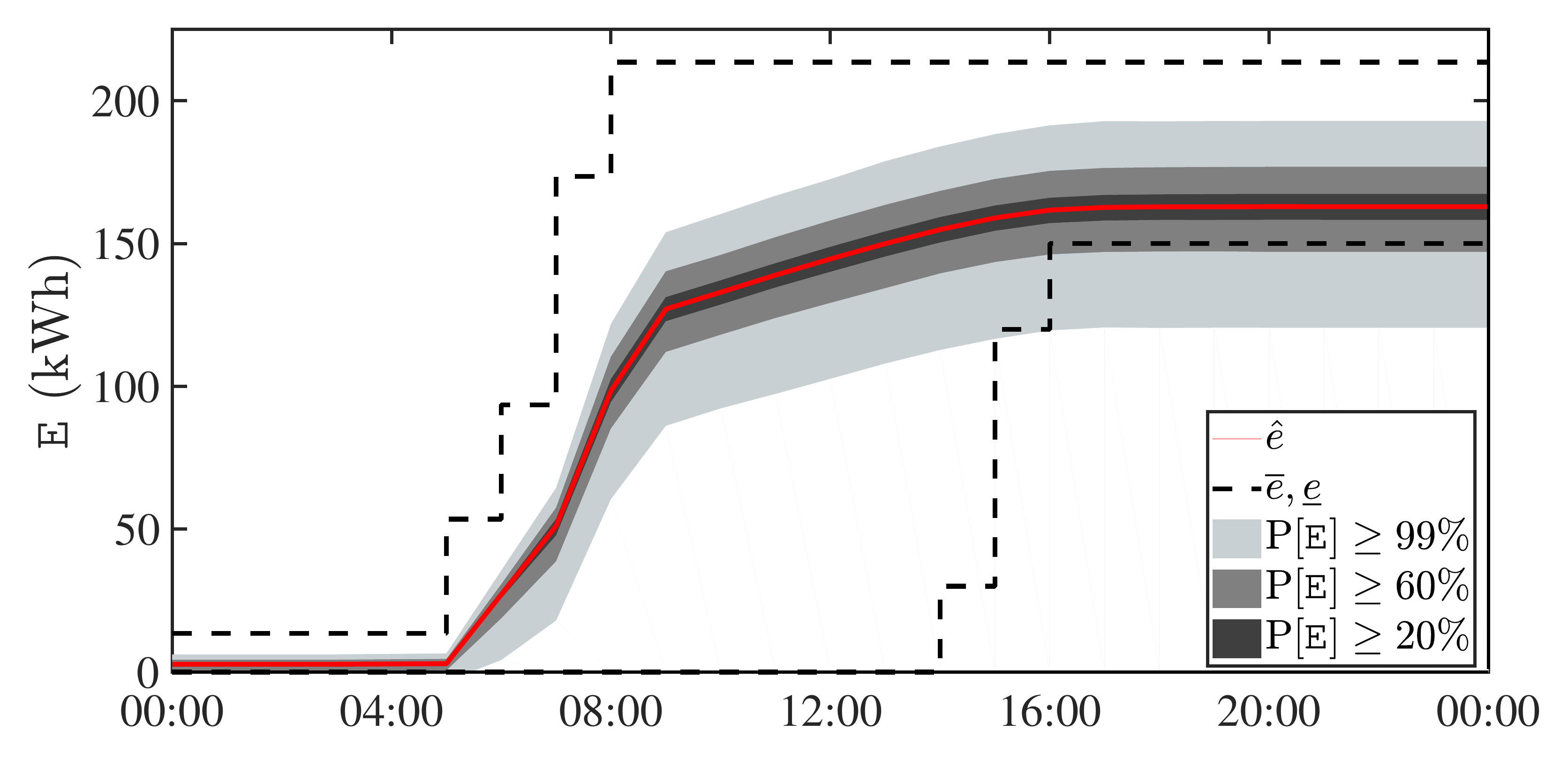}
        \vspace{-0.1cm}
    \caption{Graphical representation of the stochastic energy state $\rv{E}$ and of its corresponding constraint with $(1-\varepsilon) = 0.75$.}
    \label{fig:energy_cone}
    \vspace{-0.5cm}
\end{figure}

Indeed, the ``energy increment" $\rvevh{j}(\arrvh{j})$ is a stochastic variable affecting only the stochastic part of $\rv{E}(k)$
\begin{align*}
\rv{\Delta E}(k+1) =& \rv{\Delta E}(k) + \rv{\Delta P}(k)\Dtk + \sum_{v\in \mathcal{A}_{\KDiS}(k)} \rvevh{v}(\arrvh{v}) \nonumber \\
=& \rv{\Delta E}(k) - \rv{\Delta L}(k)\Dtk +\sum_{v\in \mathcal{A}_{\KDiS}(k)}\rvevh{v}(\arrvh{v}) \nonumber \\
=& - \sum_{i=\kstart}^{k} \rv{\Delta L}(i)\Dtk+\sum_{v\in \mathcal{A}_{\KDiS}(k)} \rvevh{v}(\arrvh{v}),
\end{align*}
with initial condition $\rv{\Delta E}(\kstart) = 0$.\footnote{The state of charge of the ``time-varying" battery for $k=\kstart$ is considered to be known at $\kzero$. Furthermore, note that the losses are neglected in the stochastic energy balance ($\mu_n=0$). See 
 \cite{Appino18a} for further details.}
In order to simplify the notation, we consider 
\begin{align*}
\rv{\Delta E}_{\text{l}}(k) = \sum_{i=\kstart}^{k-1} \rv{\Delta L}(i) \Dtk,
\end{align*}
and therefore we have
\begin{equation}
\rv{\Delta E}(k) = - \rv{\Delta E}_{\text{l}}(k) +\sum_{v\in \mathcal{A}_{\KDiS}(k)} \rvevh{v}(\arrvh{v}) \label{eq:deviation_e_bal}.
\end{equation}
Note that $\rv{\Delta P}_{\text{l}}(k)$ and $\rv{\Delta P}_{\text{l}}(k')$ with $k\not= k'$ are correlated. 
Thus, direct forecasting of the PDF $f_{\rv{\Delta E}_{\text{l}}(k)}(\Delta \eload)$ is employed to implicitly account for this correlation.
Additionally, we assume that $\rv{\Delta E}_{\text{l}}(k)$ and $\rvevh{j}$ are independent, and so are $\rvevh{j}$ and $\rvevh{h}$ with $j\not= h$.
This is a realistic assumption, as the state of charge of each PEV and the uncontrolled load/generation---on which $\rv{\Delta E}(k)$ depends---refer to completely different devices.
Therefore, $f_{\rv{\Delta E}(k)}(\Delta \eagg)$, is given by the convolution of $f_{\rv{\Delta E}_{\text{l}}(k)}(\Delta \eagg)$ and $f_{\rvevh{v}}(\evh{v})$ for each $v \in \mathcal{A}_{\KDiS}(k)$\cite{Carlton17}. 
In other words, $f_{\rv{\Delta E}(k)}(\Delta e) = \tilde{f}_{\max(\mathcal{A}_{\KDiS}(k))}^k$ with
\begin{equation}
\label{eq:convolution_ec}
\tilde{f}_j^k = \tilde{f}_{j-1}^k \ast f_{\rvevh{v}}(y), \, j \in \mathcal{A}_{\KDiS}(k) \text{ and  }\tilde{f}_{0}=f_{\rv{\Delta E}_{\text{l}}(k)}(-y), 
\end{equation}
where $\ast$ indicates the operation of convolution and the support of $f_{\rvevh{v}}(y)$ and $f_{\rv{\Delta E}(k)}(y)$ is trivially extended to the real line by setting them to zero outside of their original support.

\subsubsection{Scheduling Algorithm} 
\label{sec:scheduling_algorithm_fst}
Here, we propose computing a DiS satisfying the requirements of Section \ref{sec:req_disp_sch} via the following optimization problem:
\begin{subequations}
\label{eq:pfs_opt_problem}
\begin{align}
\min_{
\begin{subarray}{c}
  \{\mathbf{x}(k)\}_{k\in \mathcal{S}'}
  \end{subarray}} &\sum_{k \in \mathcal{S}'} \left( C\left(\Pgsch{k}\right) + \alpha_{\epsilon} \epsilon(k) \right)\\
\text{s.t. \,\,} & \Ivect \expPc{k} = \Ivect \Pgsch{k} - \Ivect \expPl{k}, \label{eq:constr_exp_pow_balance}\\ 
& \expEc{k+1} = \expEc{k} + \bmu^\top \expPc{k} \Dtk, \label{eq:constr_exp_soc} \\
& \hat{\bold{p}}(k)^\top\matrixproduct\hat{\bold{p}}(k)\leq \nu, \label{eq:constr_exp_nonconv} \\
& \Pgsch{k} - \overline{p}(k) \leq \underline{l}(k), \label{eq:constr_pow_bound_down}\\  
& - \Pgsch{k} + \underline{p}(k) \leq -\overline{l}(k) , \label{eq:constr_pow_bound_up}\\
& F_{\rv{\Delta E}(k)}\left(\expEc{k} - \overline{e}(k) \right) - F_{\rv{\Delta E}(k)}\left( \expEc{k} - \underline{e}(k) \right) \nonumber \\
&\quad + (1 - \varepsilon) \leq \epsilon(k), \label{eq:constr_en_stoc}\\
& \expEc{k+1} - \expEc{k} \leq \frac{\hat{e}(\depvh{v'}) - \underline{e}_0}{\kstart + S + S' - \depvh{v'}} \label{eq:constr_end_cond}.
\end{align}
\end{subequations}
Therein $\mathcal{S}'=[\kstart, \kstart + S + S'] \subset \mathbb{N}$ denotes the (extended) scheduling horizon, $\matrixproduct=\begin{bsmallmatrix}
0 & 1\\ 
0 & 0
\end{bsmallmatrix}$, and the decision variables are collected in ${\bold{x}(k) := [\Pgsch{k}^\top, \expPc{k}^\top, \epsilon(k)]^\top \in \mathbb{R}^5}$.

We remark that constraints \eqref{eq:constr_exp_pow_balance}-\eqref{eq:constr_en_stoc} are equivalent to those in \cite[(24)]{Appino18a}. 
Hence we only discuss them briefly.
Constraints \eqref{eq:constr_exp_pow_balance}-\eqref{eq:constr_exp_nonconv} relate to the expected values and represent the power balance \eqref{eq:exp_ess_power}, the dynamics of the expected SOC \eqref{eq:deterministic_ec-dynamics}, and the impossibility to have power flowing at the same time in and out of the ``time-varying battery". 
Specifically, \eqref{eq:constr_exp_nonconv} imposes that the product of the elements of $\pagg(k)$ remains below a given value $\nu$.
Constraints \eqref{eq:constr_pow_bound_down}-\eqref{eq:constr_en_stoc} ensure the satisfaction of the reliability requirement at each time step, i.e. that ${\text{P}[ \CP \cap \CE] \geq (1 - \varepsilon)}$, where events $\CP$ and $\CE$ corresponds to satisfaction of \eqref{eq:det_pow_limit_aggregated}-\eqref{eq:det_en_limit_aggregated}
\begin{align*}
\CP &= \left \lbrace \underline{p} \leq \Pgsch{k} - \expPl{k} - \rv{\Delta L}(k) \leq \overline{p} \right \rbrace, \nonumber \\
\CE &= \left \lbrace \underline{e} \leq \expEc{k} - \rv{\Delta E}(k) \leq \overline{e} \right\rbrace. 
\end{align*}
The parameters of \eqref{eq:constr_pow_bound_down}-\eqref{eq:constr_en_stoc} are retrieved by probabilistic forecasts. In particular, $\underline{l}(k)$ and $\overline{l}(k)$ are such that ${\text{P} \Big[ \rv{L}(k) \in [\underline{l}(k),\overline{l}(k)] \Big] \simeq 1}$ and $F_{\rv{\Delta E}(k)}\left(\Delta e\right)$ is evaluated as described in Section \ref{sec:stoch_model_energy}.
Slack variable $\epsilon(k) \in \mathbb{R}^+_0$ allows for constraint softening, with penalization weight $\alpha_{\epsilon}$ sufficiently large \cite{Kerrigan00}.
Fig. \ref{fig:energy_cone} illustrates constraint \eqref{eq:constr_en_stoc} graphically. 
Therein, the red line represents the expected state of charge $\expEc{k}$, and the areas in grey represent values that the realization $\eagg(k)$ can assume with a certain probability. 
Constraint \eqref{eq:constr_en_stoc} enforces that the probability of having $\eagg(k)$ within the energy limits of the ``time-varying battery" for $k \in \KDiS$ is at least equal to $(1-\varepsilon)$. 
Graphically, this means that the probability associated to the energy states laying within the energy limits---depicted with dashed lines---has to be at least $(1-\varepsilon)$.

Finally, constraint \eqref{eq:constr_end_cond} deals with the asymmetry between the arrival and the departure of a PEV in the model of the ``time-varying battery".
The solution of \eqref{eq:pfs_opt_problem} should avoid schedules that make use of the energy excess $\evh{v}(\depvh{v})- \evhmind{v}$ leaving with a PEV, as this energy is not physically available anymore.
Recall that the operating requirements (Section \ref{sec:operating_requirements}) demands this energy to be as small as possible.
Thus, any eventual excess of energy should be stored into the PEVs only if the ESS is fully charged, i.e. $\evh{v}(\depvh{v}) > \evhmind{v}$ only if $\evh{0}(\depvh{v}) = \overline{e}_{0}$.
Considering that this latter condition is imposed by lower control levels in compliance to the operating requirements---see Section \ref{sec:stage2}---, we additionally include constraint \eqref{eq:constr_end_cond} to \eqref{eq:pfs_opt_problem} which guarantees that---limiting the analysis to the expected values---once the last PEV has left the DCS, not more than the energy excess stored in the ESS is scheduled to be sold to the utility grid.\footnote{In case the DCS does not have an ESS, the PEV leaving last should have priority. In this case, \eqref{eq:constr_end_cond} is replaced with $\expEc{k+1} - \expEc{k} \geq 0$ for all $k \in [\depvh{v'}, \kstart + S + S']$.}
Therein, parameter $\depvh{v'}$ indicates the departure time of the last vehicle leaving the DCS, i.e. $\depvh{v'} = \max(\depvh{v} , v\in \mathcal{V}_S)$. 

\begin{rema}[Implementation aspects]
In implementation several aspects should be considered: i) relaxation of non-convex constraint \eqref{eq:constr_exp_nonconv} with sufficiently small $\nu$, ii) extension of the optimization horizon of $S'\in \mathbb{N}$, and iii) estimation of the initial state $\expEc{\kstart}$, see \cite[Remarks 1-3]{Appino18a} for further details.
\end{rema}

\subsection{Stage 2: Power Outputs of the Plug-in Electric Vehicles}
\label{sec:stage2}
The power outputs of the PEVs at $k$, $\pvh{v}(k)$ for $v \in \mathcal{V}_{\KDiS}(k)$, are determined at the second stage,
following a receding horizon approach.
The trajectory for $\pvh{v}(k)$ over a specific interval is attained by means of optimization; a new optimization is carried out at each time step based on most recent data, and the trajectories are updated accordingly.
For the sake of simplicity, we set the second-stage optimization to be performed at each step of the DiS.
In other words, $\pvh{v}(k+1)$ is computed at $k$.
The second-stage optimization solved at $k$ covers an interval $\mathcal{M}=[k+1,k+1+M]\subset\mathbb{N}$.
No aggregation is required; Fig. \ref{fig:scheduling_steps}b depicts the model of the DCS used at this stage.
The objective of the optimization is to set $\pvh{v}(h)$ and the expected power output of the ESS, $\expPs{h}$, for $h \in \mathcal{M}$ such that $\pgrid(h)$ follows the DiS, i.e.
\begin{subequations}
\label{eq:pev_sch_opt_problem}
\begin{align}
\min_{
\begin{subarray}{c}
  \{\bold{x}^M(h)\}_{h\in \mathcal{M}}
  \end{subarray}} 
  &\sum_{h \in \mathcal{M}} {\alpha}_\sigma \Ivect \dpg(h) + \sum_{\mathclap{v\in \mathcal{L}(\mathcal{M})}} c^\text{d}_v \left( \evh{v}(\depvh{v}) -  \evhmind{v} \right)^2 \label{eq:second_stage_cost_function}\\
\text{s.t. \,\,} & \Ivect \Pgsch{h} +  \Ivect \dpg(h)  = \nonumber  \\ 
&\quad \Ivect \expPl{h|k} + \Ivect \expPs{h} + \sum_{\mathclap{v\in \mathcal{V}_{\mathcal{M}}(h)}} \Ivect \pvh{v}(h), \label{eq:second_stage_pb}\\
&\text{\eqref{eq:complete_pev_din_det}, \eqref{eq:det_ess_limit} for } j \in \mathcal{V}^0_{\mathcal{K}}(k), \label{eq:second_stage_devices_con}\\
&\text{\eqref{eq:minimum_SOC_PEV} for }v \in \mathcal{L}(\mathcal{M})\label{eq:second_stage_terminal_con}\\
&\pvh{v}(h) =[0 \,\, 0]^\top  \text{ for } v \notin \mathcal{V}_{\KDiS}(h), \label{eq:second_stage_set_zero}
\end{align}
\end{subequations}
with decision variables collected in vector
\begin{align*}
 \bold{x}^M(h) := [\dpg(h)^\top, \hat{\bold{p}}_{0}(h)^\top, \pvh{1}(h)^\top, ..., \pvh{V}(h)^\top]^\top \in \mathbb{R}^{2(2+V)},
\end{align*}
and $\mathcal{L}(\mathcal{M}) = \{v \in \bigcup_{h\in \mathcal{M}} \mathcal{V}_{\KDiS}(h) |  \depvh{v} \in \mathcal{M} \}$.

Therein, \eqref{eq:second_stage_pb} represents the power balance, \eqref{eq:second_stage_devices_con} and \eqref{eq:second_stage_terminal_con} refer to the constraints of the connected devices,\footnote{Note a slight simplification of notation: in \eqref{eq:second_stage_devices_con} one should consider the \textit{expected} power output and SOC of the ESS, $\expPs{k}$ and $\hat{e}_0$, and not their realizations $\pvh{0}$ and $\evh{0}$.} and \eqref{eq:second_stage_set_zero} sets to zero the power outputs of the not-connected PEVs.
Note that constraint softening is used to guarantee the feasibility of the power balance, with slack variable $\dpg(h)$ representing unavoidable imbalances.
Therefore, the cost function in \eqref{eq:second_stage_cost_function} contains a penalization of $\dpg(h)$ with an appropriate weight ${\alpha}_\sigma$ \cite{Kerrigan00}.
Additionally, over-charging of the $v$-th PEV is also penalized with weight $c^\text{d}_v$, which can be different for different PEVs and can change in subsequent iteration of \eqref{eq:second_stage_pb}.
Practically speaking, it may be useful to set higher values of $c^\text{d}_v$ for the PEVs leaving at first, such that the ``departure" of any eventual energy excess is delayed as much as possible.
This way, any excess of energy stored in the PEVs is still usable for compensation of uncertainty. 

Note that the values of the intermittent power output over $\mathcal{M}$, i.e.  $\{\pload(h)\}_{h\in \mathcal{M}}$, and the SOC upon arrival of the not-yet-connected PEVs are still uncertain at this stage.
However, differently from the first stage, the second-stage optimization does not contain random variables, making use of \textit{deterministic} forecasts $\expPl{h|k}$ and deterministic vehicle parameters.
In fact, long-term uncertainty is already accounted for in the DiS itself and short-term uncertainty is compensated by the ESS (cf. Section \ref{sec:stage3}).

Finally, note that the DiS over $\mathcal{M}$ is a parameter of \eqref{eq:pev_sch_opt_problem}. 
Therefore, the horizon $M$ should be such that the schedule for the following day is always available before being required in \eqref{eq:pev_sch_opt_problem}, i.e. $M < (\kstart - \kzero)$. 

\subsection{Stage 3: Power Outputs of the Energy Storage System}
\label{sec:stage3}

The power output of the ESS, $\psto$, is finally computed at the third stage, once the uncontrolled power output is known. 
The system model used at this stage is depicted in Fig. \ref{fig:scheduling_steps}c.
At $k$, the power outputs of the PEVs follow the references computed at the second stage, $\pvh{v}(k)$, while the ESS is controlled such that the power exchange with the grid follows ${p_{\text{ref}}(k)= \Ivect \Pgsch{k} + \Ivect \dpg(k)}$ as much as possible, e.g. \cite[Section IV]{Sossan16a}.
In other words, the actual power output of the storage $p_0(k)$ complies with the realization $l(k)$ in accordance to the power balance
\begin{equation*}
p_0(k) = \Ivect \Pgsch{k} + \Ivect \dpg(k) - l(k) - \sum_{\mathclap{j\in \mathcal{V}_{\mathcal{S}}(k)}} \Ivect \pvh{j}(k).
\end{equation*}
 
\section{Simulations and Results}
\label{sec:sim_and_results}

The efficacy of the proposed three-stage scheduling and control scheme is assessed simulating a realistic test case built upon real data of uncontrolled generation and vehicle traffic.
The specifics of the selected test case and the simulation results are reported and discussed in the following.

\subsection{Test Case}
The test case is a small parking lot of an office building. 
The parking lot is provided with a PV generator, an ESS and five charging stations. 
All the devices are connected to the distribution grid and operated together as a DCS.
A DCS controller is responsible to: i) communicate the DiS to the system operator, ii) regulate the power output of the chargers, and iii) set the value of $p_{\text{ref}}$ for the low-level storage controller. 
The details of the various components are as follows.

\subsubsection{Photovoltaic Generator}
The PV generator has 10 kWp, and it is controlled to track its maximum power point.
The PV generation data is taken from the solar track dataset of the Global Energy Forecasting Competition of 2014 \cite{Hong16b}. This dataset consists of time series---with hourly resolution---of measured PV generation and of their corresponding solar radiation forecasts. 
The measurements have been conducted in an unspecified region of Australia.

\subsubsection{Energy Storage System}
The parameters of the ESS are retrieved from the catalog of a commercial producer.\footnote{www.tesla.com/powerwall [Accessed: 15-Jan-2018]} 
Table \ref{tab:ESS_PEV_param} reports the power and energy limits (only the usable capacity is considered), while $\mun = 0.05$.

\subsubsection{Plug-in Electric Vehicle}
\begin{figure}
	\centering
        \includegraphics[width=0.46\textwidth]{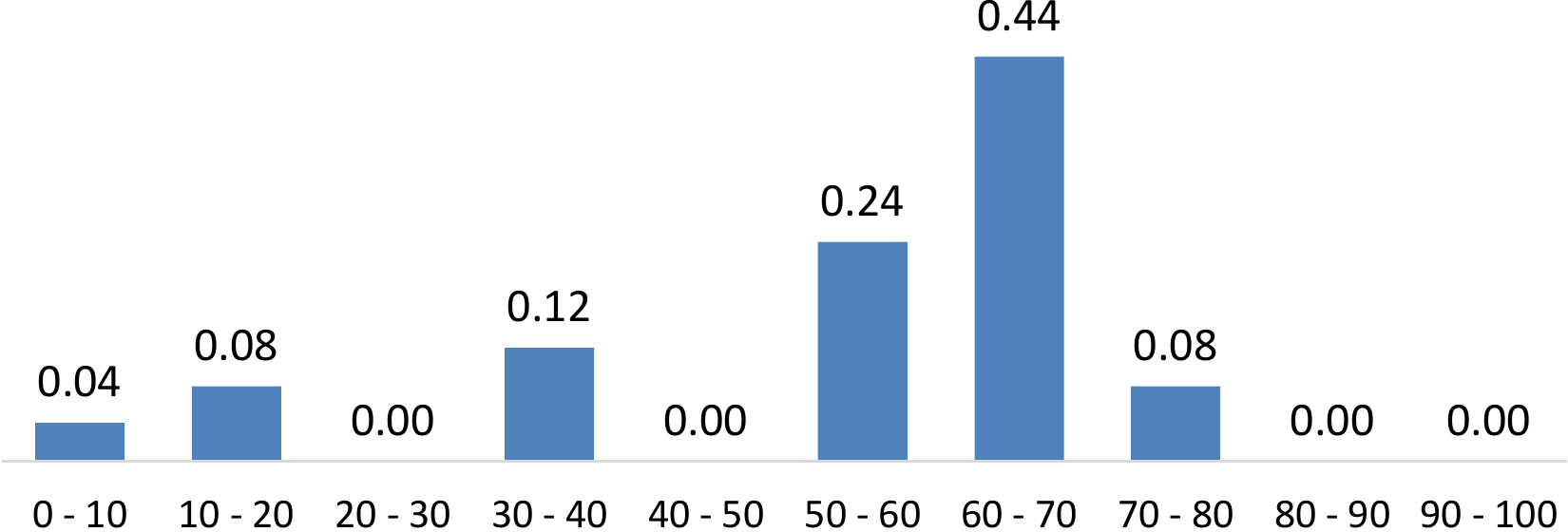}
    \caption{Relative frequency of SOC (in \%) of the PEVs upon arrival.}
    \label{fig:frequency_SoC}
    \vspace{-0.45cm}
\end{figure}
For the sake of simplicity, we consider that all the charging stations are used daily, i.e. ${\mathcal{V}_{\mathcal{S}} = [1 \hdots 5]}$ every day (weekends included),
and that all the connecting PEVs have the same characteristics. 
These parameters have been chosen after a consultation of the catalogs of different producers, as average values among the selected products.\footnote{Tesla Model 3, Chevrolet Bolt 2017, Nissan Leaf 2018, BMW i3s} 
Table \ref{tab:ESS_PEV_param} reports the power and energy limits, while $\mun = 0.05$ (same as for the ESS, cf. Section \ref{sec:system_description}).
We consider the challenging situation in which the owners of the PEVs do not allow for a vehicle-to-grid service, i.e. the DCS cannot discharge a connected PEV. 
Thus, $\underline{p}_{v}$ is set to $0$.
The minimum charge required at departure is $\evhmind{v}=30$kWh $\forall v \in \mathcal{V}_{\KDiS}$.
The data on the vehicle mobility (SOC upon arrival, arrival and departure times), have been selected on the base of a survey on vehicle usage conducted within the Institute for Automation and Applied Informatics at the Karlsruhe Institute of Technology over a time period of three months.\footnote{For space reasons, we limit to show here the results of the survey that are relevant for the present work. 
Please note that the only aim of the survey is to provide a realistic test case for the proposed scheme and does not allow to infer general trends on the vehicle usage.
However, our data is aligned with the tendencies highlighted by other works, cf. \cite{Sarabi16,Lee12}.}

The arrival times of the five PEVs are assigned proportionally to their statistical frequency.
The departure times, instead, are chosen as the weighted average of the departure times of the PEVs arriving within the same hour, rounded down.   
The resulting data is listed in Table \ref{PEV_a_and_d}.
The (statistical) relative frequency of the initial SOC is estimated on the base of the traveling distance, as in \cite{Sarabi16}. 
We assume an autonomy of $280$km with full charge and that each $v$-th PEV charges only at the DCS, up to $\evhmind{v}$.
The results are depicted in Fig. \ref{fig:frequency_SoC}.
In simulation, $\evh{v}(\arrvh{v})$ is randomly extracted in accordance with this frequency. 

\subsubsection{Forecasts}
The data of PV generation and radiation forecasts is used to train several quantile regressions based on a method described in \cite{GonzalezOrdiano17} with the open-source MATLAB
toolbox SciXMiner \cite{Mikut17}. 
Thereafter, probabilistic forecasts for both power and energy are obtained using the procedure described in \cite[Section 5.2]{Appino18a}. 
Then, we apply \eqref{eq:convolution_ec} approximating $f_{\rvevh{v}}(y)$ by the frequency illustrated in Fig. \ref{fig:frequency_SoC}. 
Finally, the analytic description of $F_{\rv{\Delta E}_{\text{c}}(k)}(\Delta e_{\text{c}})$ required in \eqref{eq:pfs_opt_problem} is obtained as in \cite[Footnote 7]{Appino18b}.

\subsubsection{Simulation Setup}
The simulations investigate six different weeks in the time frame between October 2013 and May 2014. 
Seasonal changes are considered by selecting weeks from different months.
Multiple simulations are conducted for each week, with different values for the reliability level $(1 - \varepsilon)$.
We follow the rules of day-ahead markets: the DiS covers a 24h-long time interval divided in hour-long steps, and it is computed at 12:00 of the previous day, i.e. $\Dtk=1$h, $\kstart=12$, $S=24$ and $\kzero$ at 12:00.
The coefficients of the cost functions are fixed and equal to $\bold{C}=\begin{bsmallmatrix}
0.05 & 0\\ 
0 & 0.05 
\end{bsmallmatrix}$, $\bold{c}=[0.3 \,\, 0.15]$, $c^\text{i}=0.02$, and $c^\text{d}_v=0.03$. 
Horizon $M$ is set to 11.
The simulations are implemented in MATLAB, using standard open-source optimization tools developed in the systems and control community.
Specifically, we use CasaDi \cite{Andersson13b} with IPOPT. 
All the computations have been performed using a PC with an Intel\textsuperscript{\textregistered} Core\textsuperscript{TM} i5-6400 CPU at 2.70 GHz and 8.00 GB RAM. 

\begin{table}
\renewcommand{\arraystretch}{1.4}
\caption{Parameters of ESS and PEV} \label{tab:ESS_PEV_param} 
	\vspace{-0.2cm}
\begin{center}
  \begin{tabular}{c l c l | c l c l}
  	\hline 
    \hline
    \multicolumn{4}{c}{ESS parameters} & \multicolumn{4}{c}{PEV parameters} \\
    \hline
     $\underline{p}_0$ &  -5 kW & $\overline{p}_0$ &  5 kW & $\underline{p}_{v}$ &  0 kW & $\overline{p}_{v}$ & 10 kW \\
     $\underline{e}_0$ &  0 kWh  &  $\overline{e}_0$ & 13 kWh  & $\underline{e}_{v}$ &  7 kWh & $\overline{e}_{v}$ & 40 kWh\\
    \hline
    \hline
  \end{tabular}
\end{center}
\vspace{-0.2cm}
\end{table}

\begin{table}
\renewcommand{\arraystretch}{1.3}
\caption{Arrival/departure data} \label{PEV_a_and_d} 
	\vspace{-0.2cm}
\begin{center}
  \begin{tabular}{c | c c c c c}
  	\hline
    \hline
    $v$ & 1 & 2 & 3 & 4 & 5\\
    \hline
	arrival time & 07:00 & 08:00 & 09:00 & 09:00 & 10:00 \\
	departure time	& 17:00 & 16:00 & 17:00 & 17:00 & 18:00	\\
    \hline
    \hline
  \end{tabular}
\end{center}
	\vspace{-0.5cm}
\end{table}

\subsection{Results}
\label{sec:results}

First, note that the average time to compute the DiS, reported in Table \ref{tab:simulation_results}, is always of fractions of a second.
Hence, the computational load does not appear to restrain the implementation.

\begin{table}
\small
\renewcommand{\arraystretch}{1.3}
	\caption{Simulation results
				\label{tab:simulation_results}}
	\vspace{-0.2cm}
	\begin{center}
		\begin{tabular}{ l | c c c c }
		 	\hline
    	 	\hline 
			$(1 - \varepsilon_E)$ &$\phantom{-}0.55$ &$\phantom{-}0.65$ &$\phantom{-}0.75$ &$\phantom{-}0.85$ \\
			\hline
			\hline
			$R^{\gamma}(\ds)$ & $\phantom{-}0.77$ &$\phantom{-}0.83$ &$\phantom{-}0.87$ &$\phantom{-}0.89$ \\
			balancing energy (kWh) & $\phantom{-}7.67$ &$\phantom{-}5.94$ &$\phantom{-}4.12$ &$\phantom{-}3.51$ \\
			\hline 
			computation time (s) & $\phantom{-}0.51$ &$\phantom{-}0.52$ &$\phantom{-}0.52$ &$\phantom{-}0.54$ \\
			\hline 
			cost $\ds$ (\euro) & $-0.25$ &$\phantom{-}0.37$ &$\phantom{-}1.28$ &$\phantom{-}2.86$ \\
			cost $\{\dpg(k)\}_{k \in \mathcal{\KDiS}}$ (\euro) & $\phantom{-}4.60$ &$\phantom{-}3.56$ 
				&$\phantom{-}2.47$ &$\phantom{-}2.10$ \\
			cost total (\euro) & $\phantom{-}4.35$ &$\phantom{-}3.94$ &$\phantom{-}{3.76}$ &$\phantom{-}4.96$ \\
			\hline
    		\hline
		\end{tabular}
	\end{center}
	\vspace{-0.3cm}
\end{table}

\begin{figure}
	\centering
        \includegraphics[width=0.48\textwidth]{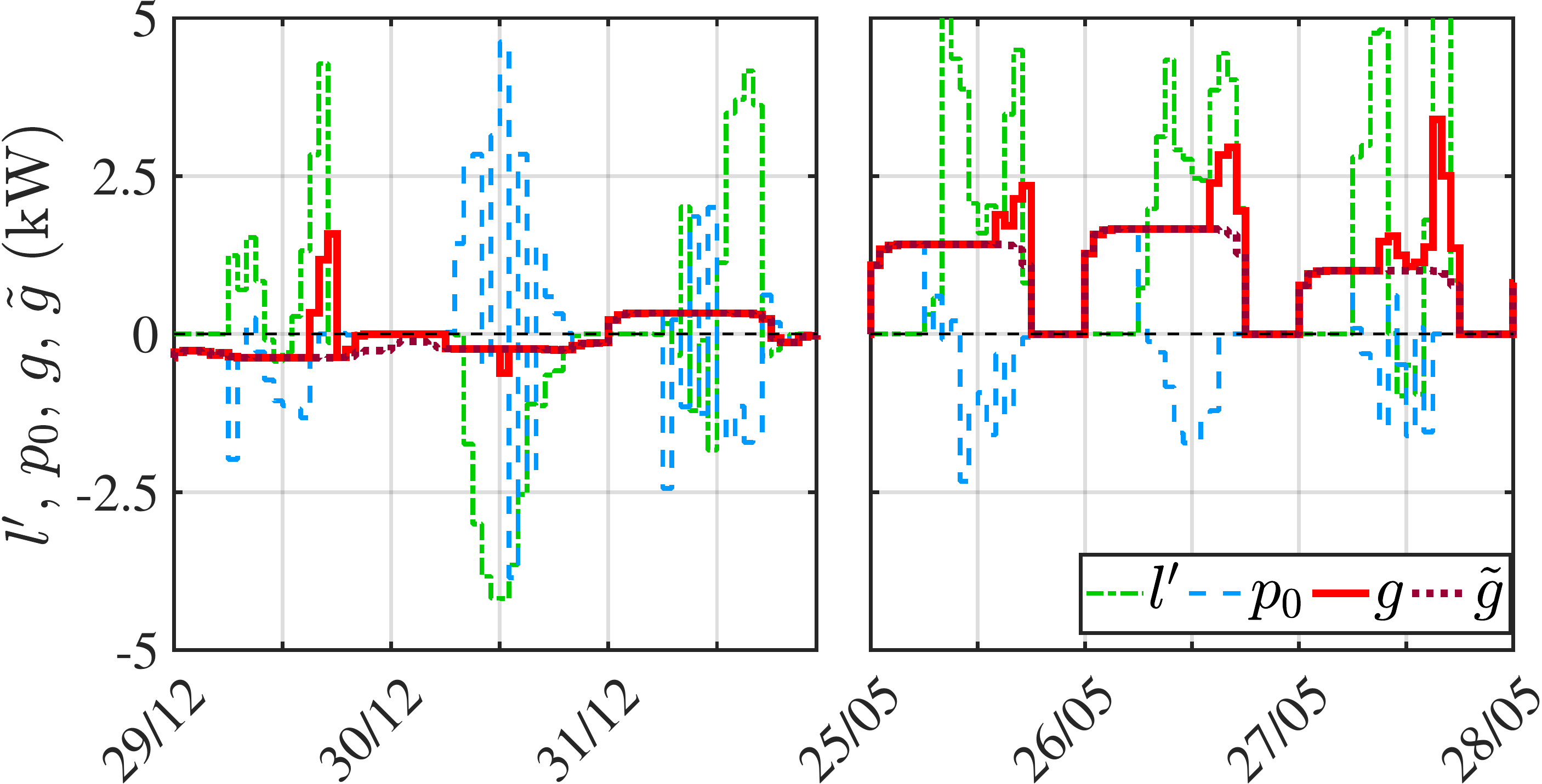}
    \caption{Power profiles over different seasons with $(1-\varepsilon) = 0.75$.}
    \label{fig:profile_dcs}
    \vspace{-0.5cm}
\end{figure}

To assess the tracking of the DiS, we define the ratio
\begin{equation*}
R^{\gamma}(\ds)=\frac{\#\left\{ k \in \KDiS \mid \left \|  \Ivect \pgrid(k) -  \Ivect \Pgsch{k} \right \| \leq \gamma \right\}}{\# \KDiS},
\end{equation*}
where $\#$ denotes the cardinality of the set and $\gamma = 10^{-4}$. 
The average values of $R^{\gamma}(\ds)$ resulting from different choices of $(1 - \varepsilon)$ are reported in Table \ref{tab:simulation_results}.
The realized $R^{\gamma}(\ds)$ is always higher than $(1 - \varepsilon)$, meaning that the proposed method meets the reliability requirement.
The accurate tracking of the DiS can be observed also in Fig. \ref{fig:profile_dcs}, showing various power profiles over simulated summer days (on the left) and fall days (on the right). 
Therein, the green dashed line represents an eventual baseline prosumption 
\begin{equation}
l'(k) = l(k) + \sum_{v\in \mathcal{V}_{\KDiS}(k)}\frac{\evhmind{v} - \evh{v}(\arrvh{v})}{\left(\depvh{v} - \arrvh{v} \right)},
\end{equation} 
corresponding to the case where the charge of each $v$-th PEVs cannot be manipulated and remains constant over the interval $[\arrvh{v}, \depvh{v}]$. 
The light blue dashed line outlines the ESS power output, $p_0$.
The dark red dotted line depicts the DiS $\tilde{g}(k)=\Ivect \Pgsch{k}$ and the red line is the realized power exchange $g(k)=\Ivect \pgrid(k)$. 
Note that the DiS is not trivial and varies daily in accordance with the forecasts and the initial SOC of the ESS. 
As expected, the DiS greatly changes with the season.
In Fig. \ref{fig:profile_dcs} it can be also observed that imbalances are more likely to appear towards the end of the day, where the uncertainty on the aggregated energy state is higher (see Fig. \ref{fig:energy_cone}).
Finally, Fig. \ref{fig:profile_dcs} illustrates that the proposed method highly reduces the ramp power required from the grid. 
Specifically, the average of the daily maximum difference among two subsequent values of $l'$ is of $2.98$ kW, which diminish to $1.34$ kW for $g$ ($55\%$ less). 

The improvements in the power profile can be also evaluated applying cost function \eqref{eq:cost_f}, considering the linear coefficients $\bold{c}$ and $c$ in \euro$/$kW, and the quadratic ones $\bold{C}$ in \euro$/$kW$^2$.
The hypothetical average daily cost of the baseline load $l'$, evaluated according to \eqref{eq:cost_f} as if it was dispatched, is of $7.07$\euro.
The average daily costs of the DiS, reported in Table \ref{tab:simulation_results}, are between $-0.25$\euro and $2.86$\euro. 
Table \ref{tab:simulation_results} also details an hypothetical cost of $\dpg$, with tariffs that are twice as high as the one of the DiS counting both power excess and shortage as purchased power.
From the listed results, it can be inferred that enhancing the internal reserves increases the cost of the DiS, while reducing the cost of imbalances. 
These results are aligned with what is observed in \cite{Appino18b}.

\begin{figure}
	\centering
        \includegraphics[width=0.495\textwidth]{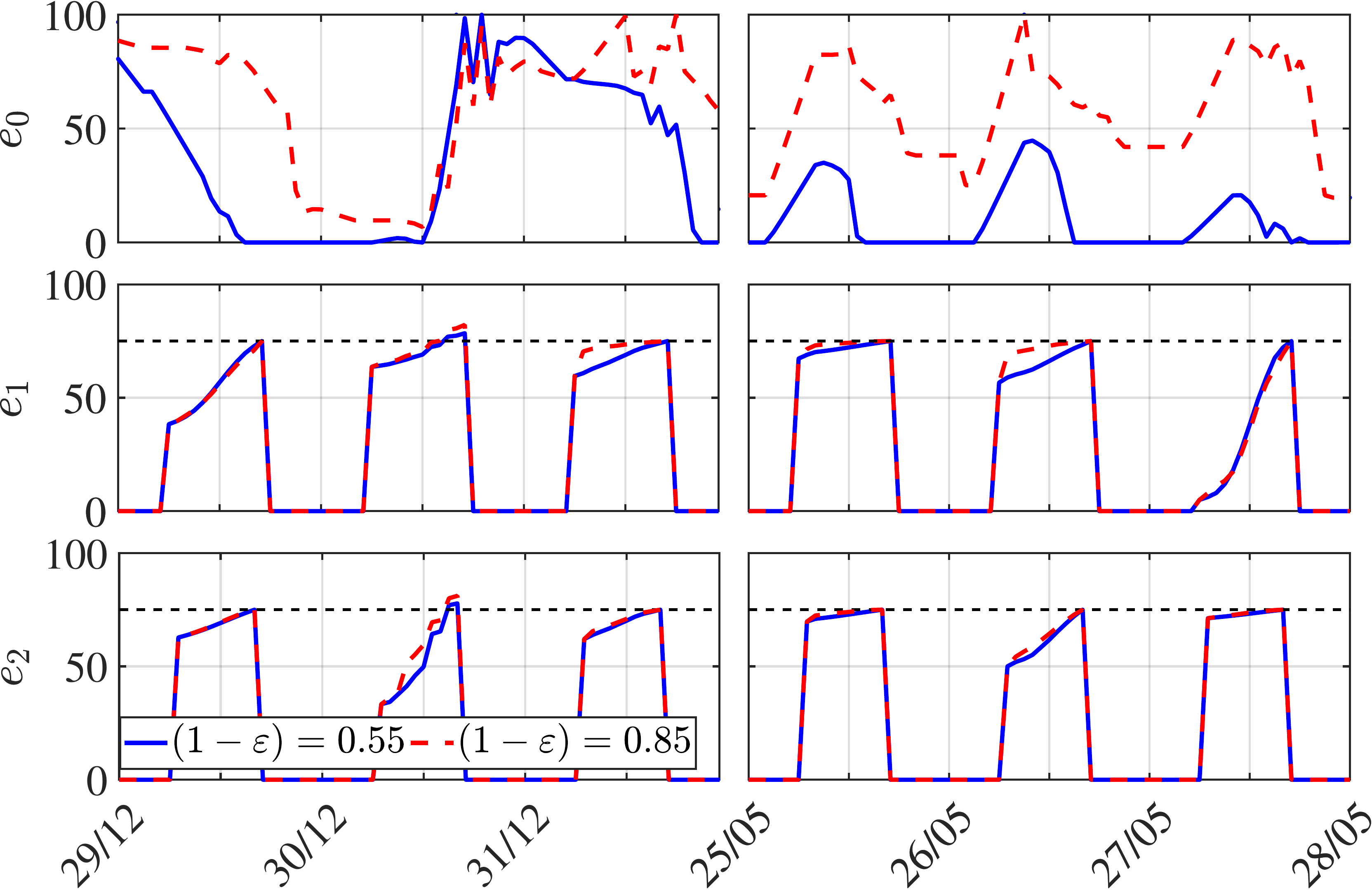}
    \caption{SOC of the ESS and two PEVs with different $(1-\varepsilon)$.}
    \label{fig:soc_ess_pev}
    \vspace{-0.5cm}
\end{figure}

Fig. \ref{fig:soc_ess_pev} depicts the SOCs of the ESS and two of the five PEVs over the same days illustrated in Fig. \ref{fig:profile_dcs}, with varying values for $(1 -\varepsilon)$.
First, note that the requirement on the minimum SOC of each PEV upon departure is always met.
Then, observe how the behavior of the system changes with $(1 - \varepsilon)$.
This is due to a different allocation of the energy reserves: increasing $(1 - \varepsilon)$ reduces the reserves required on the grid side (see Table \ref{tab:simulation_results})\footnote{We denote the average amount of energy needed each day to compensate for deviations from the DiS as ``balancing energy", which is considered positive regardless of whether it is absorbed or injected into the grid.} and increases the internal ones. 
Consequently, with low values of $(1 -\varepsilon)$: i) the available capacity of the ESS is used both to provide energy reserves and to optimize the DiS, and ii) the final SOC of each PEVs rarely exceed its respective $\evhmind{v}$.
On the other end, with high values of $(1 -\varepsilon)$: i) the ESS is almost solely used for reserves purposes, and ii) the PEVs are generally charged over their $\evhmind{v}$.
In this case the ESS is maintained charged such that it can inject power in case of underproduction and concurrently the PEVs absorb production excess.
Finally, it is possible to observe that the volatility of $l$ and of the forecasts $\expPl{k}$ (particularly evident around midday during the summer days) reflects in rapid fluctuations of $p_0$ and, consequently, of $e_0$. 

\section{Conclusions}
\label{sec:conclusion}

The present paper examined scheduling and operation of a dispatchable charging station, where dispatchability of intermittent generation is achieved compensating the uncertainties locally by means of an energy storage system and controlled charging of plug-in electric vehicles.
Our main contribution is the application of probabilistic forecasts to a dynamic aggregation of diverse devices (which results in a ``time-varying battery"), achieved by means of a three-stage approach.
Therein, we account for the uncertainty affecting both the forecasts of intermittent generation and the initial state of charge of the vehicles upon arrival.
The outcome is a dispatch schedule with a given reliability level, i.e. a probability of being realized in operation. 
The simulation results show that the requirements on vehicle charging and power dispatch are always met, thus attesting the validity of the proposed method.    
Future work will investigate the consideration of network constraints for devices which are not all connected to the same bus.


%

%



\ifCLASSOPTIONcaptionsoff
  \newpage
\fi



\bibliographystyle{IEEEtran}
\bibliography{Appino_bib}{}

\begin{thebibliography}{10}
\providecommand{\url}[1]{#1}
\csname url@samestyle\endcsname
\providecommand{\newblock}{\relax}
\providecommand{\bibinfo}[2]{#2}
\providecommand{\BIBentrySTDinterwordspacing}{\spaceskip=0pt\relax}
\providecommand{\BIBentryALTinterwordstretchfactor}{4}
\providecommand{\BIBentryALTinterwordspacing}{\spaceskip=\fontdimen2\font plus
\BIBentryALTinterwordstretchfactor\fontdimen3\font minus
  \fontdimen4\font\relax}
\providecommand{\BIBforeignlanguage}[2]{{%
\expandafter\ifx\csname l@#1\endcsname\relax
\typeout{** WARNING: IEEEtran.bst: No hyphenation pattern has been}%
\typeout{** loaded for the language `#1'. Using the pattern for}%
\typeout{** the default language instead.}%
\else
\language=\csname l@#1\endcsname
\fi
#2}}
\providecommand{\BIBdecl}{\relax}
\BIBdecl

\bibitem{Denholm07}
P.~Denholm and R.~M. Margolis, ``Evaluating the limits of solar photovoltaics
  ({PV}) in traditional electric power systems,'' \emph{Energy Policy},
  vol.~35, no.~5, pp. 2852--2861, 2007.

\bibitem{Kempton05}
W.~Kempton and J.~Tomi{\'c}, ``Vehicle-to-grid power implementation: From
  stabilizing the grid to supporting large-scale renewable energy,''
  \emph{Journal of Power Sources}, vol. 144, no.~1, pp. 280--294, 2005.

\bibitem{Mukherjee15}
J.~C. Mukherjee and A.~Gupta, ``A review of charge scheduling of electric
  vehicles in smart grid,'' \emph{IEEE Systems Journal}, vol.~9, no.~4, pp.
  1541--1553, 2015.

\bibitem{Momber15}
I.~Momber, A.~Siddiqui, T.~G. San~Rom{\'a}n, and L.~S{\"o}der, ``Risk averse
  scheduling by a pev aggregator under uncertainty,'' \emph{IEEE Trans. Power
  Syst.}, vol.~30, no.~2, pp. 882--891, 2015.

\bibitem{Kou16}
P.~Kou, D.~Liang, L.~Gao, and F.~Gao, ``Stochastic coordination of plug-in
  electric vehicles and wind turbines in microgrid: A model predictive control
  approach,'' \emph{IEEE Trans. Smart Grid}, vol.~7, no.~3, pp. 1537--1551,
  2016.

\bibitem{Appino18b}
R.~R. Appino, J.~{\'A}.~G. Ordiano, R.~Mikut, V.~Hagenmeyer, and T.~Faulwasser,
  ``Storage scheduling with stochastic uncertainties: Feasibility and cost of
  imbalances,'' in \emph{Power Systems Computation Conference (PSCC),
  2018}.\hskip 1em plus 0.5em minus 0.4em\relax IEEE, 2018, arXiv:1805.02525.

\bibitem{Mathieu13}
J.~L. Mathieu, M.~G. Vay{\'a}, and G.~Andersson, ``Uncertainty in the
  flexibility of aggregations of demand response resources,'' in
  \emph{Industrial Electronics Society, IECON 2013-39th Annual Conference of
  the IEEE}.\hskip 1em plus 0.5em minus 0.4em\relax IEEE, 2013, pp. 8052--8057.

\bibitem{Bernstein15a}
A.~Bernstein, L.~Reyes-Chamorro, J.-Y. Le~Boudec, and M.~Paolone, ``A
  composable method for real-time control of active distribution networks with
  explicit power setpoints. part i: Framework,'' \emph{Electric Power Systems
  Research}, vol. 125, pp. 254--264, 2015.

\bibitem{Subramanian13}
A.~Subramanian, M.~J. Garcia, D.~S. Callaway, K.~Poolla, and P.~Varaiya,
  ``Real-time scheduling of distributed resources,'' \emph{IEEE Trans. Smart
  Grid}, vol.~4, no.~4, pp. 2122--2130, 2013.

\bibitem{Appino18a}
R.~R. Appino, J.~{\'A}.~G. Ordiano, R.~Mikut, T.~Faulwasser, and V.~Hagenmeyer,
  ``On the use of probabilistic forecasts in scheduling of renewable energy
  sources coupled to storages,'' \emph{Applied Energy}, vol. 210, no.
  Supplement C, pp. 1207 -- 1218, 2018.

\bibitem{Sossan16a}
F.~Sossan, E.~Namor, R.~Cherkaoui, and M.~Paolone, ``Achieving the
  dispatchability of distribution feeders through prosumers data driven
  forecasting and model predictive control of electrochemical storage,''
  \emph{IEEE Trans. Sustain. Energy}, vol.~7, no.~4, pp. 1762--1777, 2016.

\bibitem{Vandael13}
S.~Vandael, B.~Claessens, M.~Hommelberg, T.~Holvoet, and G.~Deconinck, ``A
  scalable three-step approach for demand side management of plug-in hybrid
  vehicles,'' \emph{IEEE Trans. Smart Grid}, vol.~4, no.~2, pp. 720--728, 2013.

\bibitem{Wenzel17}
G.~Wenzel, M.~Negrete-Pincetic, D.~E. Olivares, J.~MacDonald, and D.~S.
  Callaway, ``Real-time charging strategies for an electric vehicle aggregator
  to provide ancillary services,'' \emph{IEEE Trans. Smart Grid}, vol.~PP,
  no.~99, 2017.

\bibitem{GonzalezOrdiano17}
J.~{\'A}. Gonz{\'a}lez~Ordiano, W.~Doneit, S.~Waczowicz, L.~Gr{\"o}ll,
  R.~Mikut, and V.~Hagenmeyer, ``Nearest-neighbor based non-parametric
  probabilistic forecasting with applications in photovoltaic systems,''
  \emph{arXiv preprint arXiv:1701.06463}, 2017.

\bibitem{Borkowska74}
B.~Borkowska, ``Probabilistic load flow,'' \emph{IEEE Trans. Power App. Syst.},
  vol. PAS-93, no.~3, pp. 752--759, 1974.

\bibitem{Lee12}
T.-K. Lee, Z.~Bareket, T.~Gordon, and Z.~S. Filipi, ``Stochastic modeling for
  studies of real-world phev usage: Driving schedule and daily temporal
  distributions,'' \emph{IEEE Trans. Veh. Technol.}, vol.~61, no.~4, pp.
  1493--1502, 2012.

\bibitem{Sarabi16}
S.~Sarabi, A.~Davigny, V.~Courtecuisse, Y.~Riffonneau, and B.~Robyns,
  ``Potential of vehicle-to-grid ancillary services considering the
  uncertainties in plug-in electric vehicle availability and
  service/localization limitations in distribution grids,'' \emph{Applied
  Energy}, vol. 171, pp. 523--540, 2016.

\bibitem{Carlton17}
M.~A. Carlton, J.~L. Devore \emph{et~al.}, \emph{Probability with Applications
  in Engineering, Science, and Technology}.\hskip 1em plus 0.5em minus
  0.4em\relax Springer, 2017.

\bibitem{Kerrigan00}
E.~C. Kerrigan and J.~M. Maciejowski, ``Soft constraints and exact penalty
  functions in model predictive control,'' in \emph{Control 2000 Conference,
  Cambridge}, 2000.

\bibitem{Hong16b}
T.~Hong, P.~Pinson, S.~Fan, H.~Zareipour, A.~Troccoli, and R.~J. Hyndman,
  ``Probabilistic energy forecasting: Global energy forecasting competition
  2014 and beyond,'' \emph{International Journal of Forecasting}, vol.~32,
  no.~3, pp. 896 -- 913, 2016.

\bibitem{Mikut17}
R.~Mikut, A.~Bartschat, W.~Doneit, J.~\'{A}ngel Gonz\'{a}lez~Ordiano,
  B.~Schott, J.~Stegmaier, S.~Waczowicz, and M.~Reischl, ``The {MATLAB} toolbox
  {SciXMiner}: User's manual and programmer's guide,'' arXiv:1704.03298, Tech.
  Rep., 2017.

\bibitem{Andersson13b}
J.~Andersson, ``{A} {G}eneral-{P}urpose {S}oftware {F}ramework for {D}ynamic
  {O}ptimization,'' {P}h{D} thesis, Arenberg Doctoral School, KU Leuven,
  October 2013.

\end{thebibliography}
\end{document}